\documentclass[letterpaper]{jpconf}
\usepackage{graphicx}
\def\deg{^\circ}
\def\gtorder{\mathrel{\raise.3ex\hbox{$>$}\mkern-14mu
             \lower0.6ex\hbox{$\sim$}}}
\def\ltorder{\mathrel{\raise.3ex\hbox{$<$}\mkern-14mu
             \lower0.6ex\hbox{$\sim$}}}

\begin{document}

\title{New Measurements of the EMC Effect in Few-Body Nuclei}

\author{J. Arrington}

\address{Physics Division, Argonne National Laboratory, Argonne, IL 60639}

\smallskip
\hskip 0.8in
\small
For the JLab E03-103 collaboration.
\normalsize

\ead{johna@anl.gov}

\begin{abstract}
Measurements of the EMC effect show that the quark distributions in nuclei are
not simply the sum of the quark distributions of the constituent nucleons.
However, interpretation of the EMC effect is limited by the lack of a reliable
baseline calculation of the effects of Fermi motion and nucleon binding.  We
present preliminary results from JLab experiment E03-103, a precise
measurement of the EMC effect in few-body and heavy nuclei.  These data
emphasize the large-$x$ region, where binding and Fermi motion effects
dominate, and thus will provide much better constraints on the effects of
binding.  These data will also allow for comparisons to calculations for
few-body nuclei, where the uncertainty in the nuclear structure is minimized.
\end{abstract}

\section{Introduction}
One topic of great interest in the study of Quantum Chromodynamics (QCD) is
the question of whether or not the internal structure of nucleons is modified
inside of the nuclear medium.  Measurements of the ``EMC effect'' provide an
unambiguous experimental indication that the quark distribution of a nucleus
is not the naive sum of the quark distributions of the individual nucleons. 
However, there have been many explanations proposed for the EMC effect, some
of which require a modification to the in-medium nucleon structure, and some
of which do not.  So while the experimental signature is clear, the
interpretation of the effect is, at present, ambiguous.

One important issue in understanding the modification of the nuclear quark
distributions at large Bjorken $x$, i.e. large quark momentum, is that binding
and the Fermi motion of the nucleons must clearly play a significant role.
While these contributions are usually discussed when describing the EMC effect
at very large $x$ values, binding leads to a modification of the distributions
at all $x$ values.  Therefore, one must have a precise and quantitative
understanding of these effects before one can determine from the data if
additional, more exotic explanations must be invoked to describe the data.

The bulk of the measurements of nuclear parton distributions are for heavy
nuclei, $A \gtorder 12$, where there are uncertainties in the details of the
nuclear structure that go into determining the effect of binding.  In
addition, there is very little data above $x=0.8$, where binding effects
should dominate.  This is the region where high precision data could be used
to evaluate calculations of binding, and yet the data in this region is of
limited precision.  Because of the lack of data to constrain the effects of
binding and the limited data for few-body nuclei, where nuclear structure
uncertainties are minimized, many calculations of the EMC effect are performed
for nuclear matter, and extrapolated to lower density when comparing to the
nuclear parton distributions.  In this case, it is difficult to be certain
that the ``traditional'' effects of binding and Fermi motion are modeled well
enough to examine the effect of more exotic effects, such as the contributions
of nuclear pions or modification of the nucleon structure.

\begin{figure}[ht]
\begin{center}
\includegraphics[height=4.2in,angle=270]{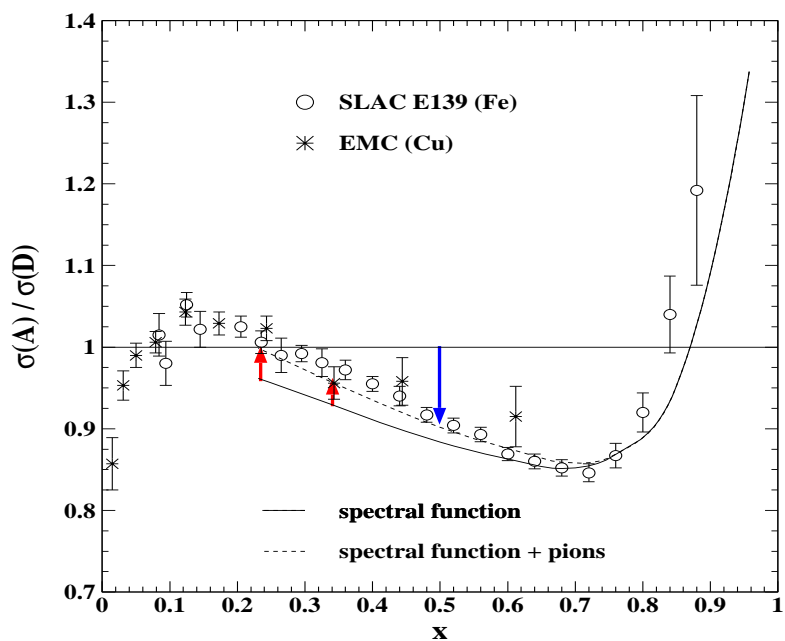}
\caption{Existing data for the EMC effect for nuclei near
Fe~\cite{ashman88,gomez94}, along with two calculations by Benhar,
Pandharipande and Sick~\cite{benhar99b}.  The solid line is their binding-only
calculation, while the dotted line includes their calculation of the
contribution from nuclear pions.}
\label{fig:calc}
\end{center}
\end{figure}

Figure~\ref{fig:calc} shows one example of a detailed binding
calculation~\cite{benhar99b}. The solid line indicates the result including
only binding, and based on this calculation, one expects that any additional
exotic effects must lead to a small enhancement of the parton distributions
for $x \ltorder 0.5$, as indicated by the red arrows.  In
Ref.~\cite{benhar99b}, the authors include the contributions of so-called
``nuclear pions'' to explain the additional enhancement needed at small $x$
values. This is in contrast to models that use a more exotic explanation to
explain the full EMC effect, including all of the suppression of the quark
distributions for the larger $x$ values, as indicated by the blue arrow.

Note that the binding calculation in Fig.~\ref{fig:calc} does not reproduce
even the somewhat low precision data at very large $x$, and therefore is not
fully explaining the effects of binding.  More precise data in this region,
especially for light nuclei where the uncertainties in the nuclear structure
are smaller, will allow for precise tests of the binding calculations that
serve as a necessary baseline when looking for additional contributions.

Previous measurements were limited at large $x$ values because of the
requirement that the data be taken in the deep inelastic scattering (DIS)
regime.  The requirement that data be limited to large $W^2$, where $W$ is the
invariant mass of the unobserved hadronic system, meant that the data had to
be taken at very large $Q^2$ for large $x$ values.  For the usual DIS
condition, $W^2 > 4$~GeV$^2$, $Q^2$ must be above 18~GeV$^2$ for $x=0.85$,
limiting the range of the measurements as the cross sections fall rapidly with
both $x$ and $Q^2$.  Recent measurements of inclusive scattering from
nuclei~\cite{filippone92, arrington01} indicate that the scaling of the nuclear
structure function expected in the DIS regime extends to lower $W$ values. By
relaxing the constraint on $W^2$, large $x$ data can be accessed at
significantly lower $Q^2$, and thus larger $x$ values can be accessed.  For
$x=0.85$, the minimum $Q^2$ is reduced to 12~(9)~GeV$^2$ for $W^2 >
3$~(2.5)~GeV$^2$.  Limited measurements of the EMC effect at lower $W^2$
suggest that the scaling continues to even lower $W^2$ and $Q^2$
values~\cite{arrington06a}.

\section{Experimental details}

Jefferson Lab (JLab) experiment E03-103~\cite{e03103} took data on few-body
and heavy nuclei, with an emphasis on the large-$x$ region, with the goal of
providing high precision data that can be used to evaluate detailed
calculations of the effects of binding.  Such calculations can then provide a
much more reliable baseline when evaluating additional contributions to the
nuclear parton distributions at lower $x$ values.

The experiment was run in Hall C at Jefferson Lab during the summer and fall
of 2004. The bulk of the data were taken at a beam energy of 5.8~GeV, with
beam currents ranging from 30 to 80 $\mu$A.  Data were taken on cryogenic
$^1$H, $^2$H, $^3$He, and $^4$He, as well as solid Be, C, Al, Cu, and Au
targets.  For the main measurements, scattered electrons were detected in the
High Momentum Spectrometer (HMS).  The HMS has a solid angle of approximately
6.8 msr, momentum acceptance of roughly $\pm$9\%, and can detect scattered
electrons with momenta up to the beam energy.  Electrons are separated from
pions using cuts on the gas Cerenkov and lead-glass shower counter, which
leave a negligible contribution of pions.

\begin{figure}[ht]
\begin{center}
\includegraphics[width=4.2in,angle=270]{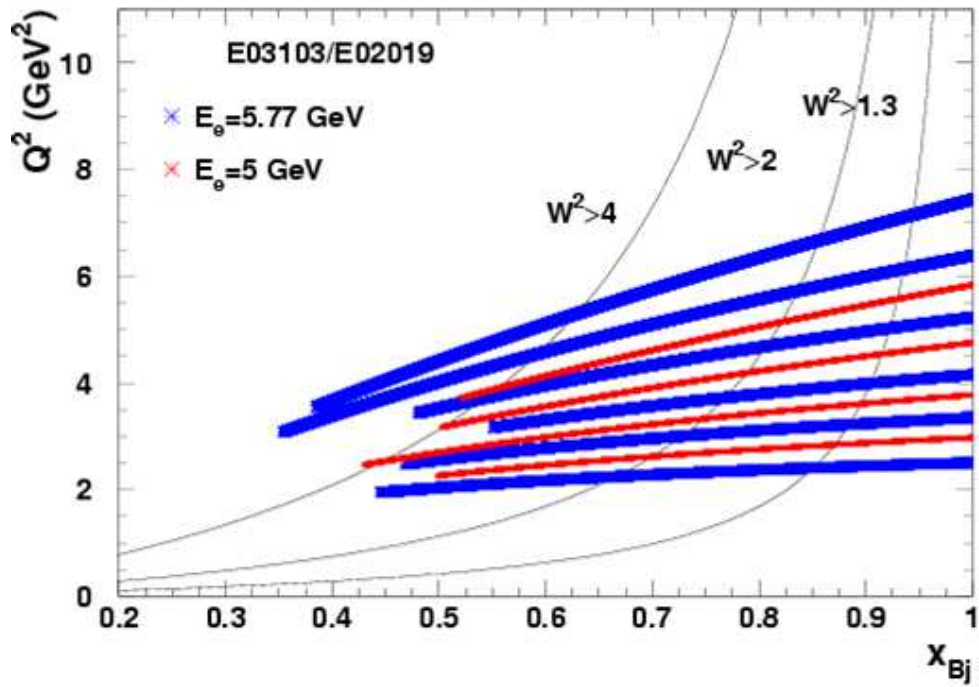}
\caption{The kinematics for the deuterium and carbon running for E03-103.
Other targets were taken only at the two highest $Q^2$ settings.  The
results from Ref.~\cite{arrington06a} were taken between $Q^2=3$~GeV$^2$ at
$x=0.6$ and $Q^2=4$~GeV$^2$ at $x=0.9$.
\label{fig:kine}}
\end{center}
\end{figure}

Data for all targets was taken at 40$\deg$ and 50$\deg$ electron scattering
angle, corresponding to $Q^2$ values near 3~GeV$^2$ at $x = 0.3$ and 6~GeV$^2$
at $x=0.9$.  Because some of this data is below the typical DIS limit
($W^2>4$~GeV$^2$), we took additional data on the $Q^2$ dependence for Carbon
and Deuterium, covering six angles between 18$\deg$ and 50$\deg$ at a beam
energy of 5.8~GeV, and four additional angles at 5.0~GeV.  This allows us to
verify that the structure function shows precise scaling in this region, as
suggested by previous data~\cite{filippone92, arrington01, arrington06a}, as
well as directly verifying that the extracted EMC ratios are independent of
$Q^2$.  Figure~\ref{fig:kine} shows the kinematics for the measurement.
Carbon and deuterium data were taken at all settings shown, while the other
targets were measured only for the two largest $Q^2$ settings.

The high luminosity and large acceptance spectrometers in Hall C, coupled with
high density cryogenic $^3$He and $^4$He targets, allowed high precision
measurements of the target ratios for both the few body nuclei and the large
$x$ region.  The largest limitation is the fact that the high $Q^2$ data is
taken at large scattering angle, due to the beam energy. This leads to some
corrections being larger than for previous measurements at higher energy. 
First, one detects both scattered electrons and electrons that come from the
production of positron-electron pairs.  These secondary electrons cannot be
separated from the beam electrons that scatter in the target.  The
contribution from this charge-symmetric background (CSB) can be quite large
for the large scattering angle, especially for very low $x$ values and the
high-$Z$ targets. Data was taken for both positive and negative polarity,
allowing a direct measurement of the CSB. For the worst case conditions, the
heavy targets (Cu and Au) at 50$\deg$ and large energy transfer (small $x$),
the contribution from the charge-symmetric background can be as large as the
signal from scattered electrons.  However, this background is a factor of
three lower for lighter nuclei, and also decreases rapidly with scattering
angle, and so is an order of magnitude smaller for the lighter nuclei at
40$\deg$.  Comparisons of the 40$\deg$ and 50$\deg$ data on Au indicate that
the uncertainty due to the subtraction on the CSB is reasonably small
($<$5\%), even when the correction itself is 100\%.  Thus this is a small
contribution to the uncertainty for the majority of the data, especially for
the few-body nuclei.

In addition, the data at large scattering angle ends up having non-trivial
corrections due to Coulomb distortion.  These corrections are estimated to be
up to 10\% for the Au data at 50$\deg$ and very small x values, according to
the updated effective momentum approximation prescription of
Ref.~\cite{aste05}.  Again, these corrections are much smaller for lighter
nuclei and for smaller scattering angles, and so again will have a small
contribution to the final uncertainty, except for the worst case conditions
mentioned above.  Finally, radiative corrections can be quite large for the
small $x$ data at large scattering angles, and this correction is sensitive to
the model of the quasielastic cross section at low $Q^2$. Detailed tests of
the model dependence are underway, and this is not expected to be a limiting
factor for any of the kinematics.

There are two main advantages to this experiment, compared to previous
measurements of the EMC effect.  First, the high density cryogenic Helium
targets allowed for much higher luminosity than available with the internal
$^3$He target at HERMES~\cite{ackerstaff99}.  The luminosity was also
noticeably larger than for the $^4$He data from SLAC E139~\cite{gomez94} and
with much smaller target density fluctuations due to beam heating.  Second, by
taking data at $W^2$ values slightly below the usual DIS cut of 4~GeV$^2$, we
were able to take data at large $x$ with much higher statistics. Previous
measurements have observed that the nuclear structure functions show extended
scaling in the Nachtmann variable $\xi$ well below the usual DIS
region~\cite{filippone92, arrington96, arrington01}.  It was also shown that
in the target ratios, the results were in agreement with the DIS measurements
even for $Q^2 \approx 3$--4~GeV$^2$ and $W^2$ values from
1.3--2.8~GeV$^2$~\cite{arrington06a}.

This can be understood as a consequence of quark-hadron duality in the
unpolarized nucleon structure functions~\cite{niculescu00b, melnitchouk05}. It
was observed originally by Bloom and Gilman~\cite{bloom70} that while the
proton inclusive structure function in the resonance region shows structure
due to the individual resonance contributions, the structure function is, on
average, in agreement with the DIS limit.  Thus, integrating over the region
of a prominent resonance yields the same result as integrating over the
equivalent region in the DIS limit.  In a nucleus, the Fermi motion provides
this averaging, so rather than seeing a duality between the resonance region
structure function when averaged over a range in $W^2$, we see that the
structure function in the resonance region is identical to the DIS limit,
after taking into account QCD evolution and target mass corrections.  In these
and other previous studies of duality in inclusive electron scattering,  the
target mass corrections were often approximated by taking the data as a
function of Nachtmann $\xi$, or some other variable, rather than Bjorken-$x$. 
While this approximation works well, duality studies have also been performed
with explicit target mass corrections~\cite{liang05, melnitchouk05,
steffens06}.

\section{Preliminary results}

We have extracted preliminary cross sections and EMC ratios for all of the
targets, but will focus here on the light nuclei, where the charge-symmetric
background and Coulomb distortion corrections are much smaller.  As discussed
in the previous section, we expect that the data should show scaling in $\xi =
2x / (1 + \sqrt{1 + 4 m_p^2 x^2/Q^2})$, even though some of the data is below
$W^2=4$~GeV$^2$.  Note that the difference between $x$ and $\xi$ goes like
$x^2/Q^2$, and so for small $x$ or very large $Q^2$, $\xi \rightarrow x$. 
Thus, for the very high energy measurements focussed on low $x$, the result is
identical for $x$ or $\xi$.  For large $x$ values, there can be a noticeable
difference between $x$ and $\xi$. In the target ratios, this is simply a shift
from $x$ to $\xi$, and so only the shape of the EMC effect will change when
plotted as a function of $\xi$ rather than $x$.  This shift is nearly
identical for both this measurement and the SLAC E139 result, shifting data at
$x=0.8$ to $\xi \approx 0.7$, while having a small effect for $x<0.5$.  As
these are the only measurements with significant data at large $x$, we show
our results as a function of $x$, rather than $\xi$, so that they can be
directly compared to previously extracted EMC ratios and calculations.

\begin{figure}[ht]
\begin{center}
\includegraphics[width=5.2in]{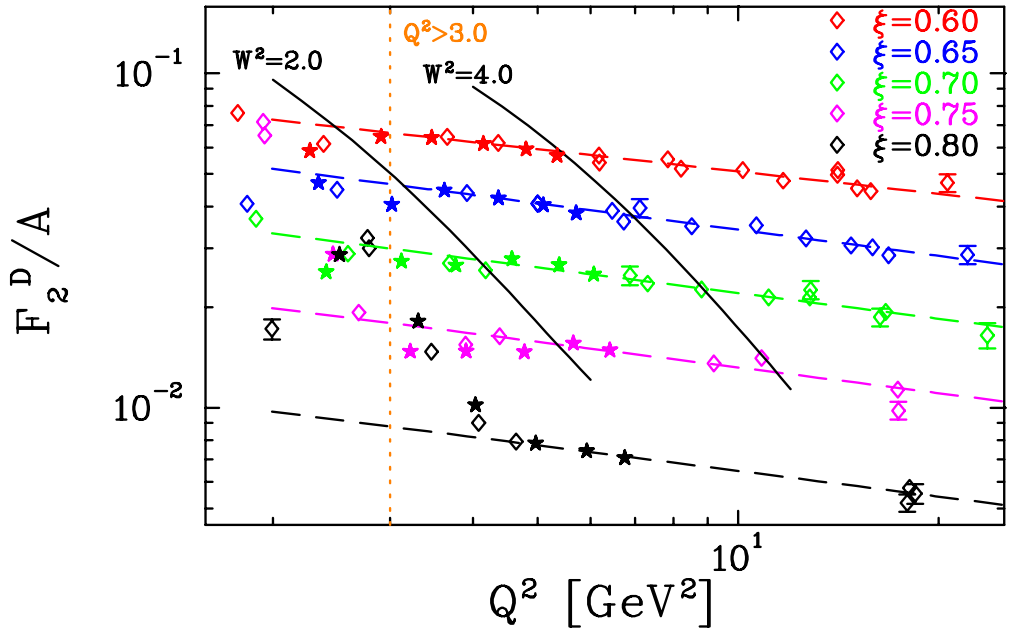}
\caption{The $Q^2$ dependence of the deuteron structure function at fixed
$\xi$ values.  The hollow points are previous SLAC and JLab data,
while the solid stars are the preliminary E03-103 results.  The
dashed lines correspond to fixed values of $d\ln(F_2)/d\ln(Q^2)$, fit to the
large $Q^2$ SLAC data.  The structure function is consistent with this 
evolution well below the DIS limit, down to $W^2=2$ for the $\xi$ values shown
here.
\label{fig:qsqf2}}
\end{center}
\end{figure}

The measurements are in the DIS region up to $x \approx 0.65$.  For larger
$x$ values, we rely on the extended scaling in nuclei.  We observe that the
structure functions for carbon and deuterium are consistent with the prediction
of QCD evolution at better than the 5\% percent level over most of the range of
the data, including over most of the resonance region at the higher $Q^2$
values from this measurement (see Fig.~\ref{fig:qsqf2}).  While we do not
see deviations from QCD scaling in the individual structure functions, any
deviations that are too small to see will be suppressed to the extent that
they are $A$-independent.

\begin{figure}[ht]
\begin{center}
\includegraphics[width=4.9in]{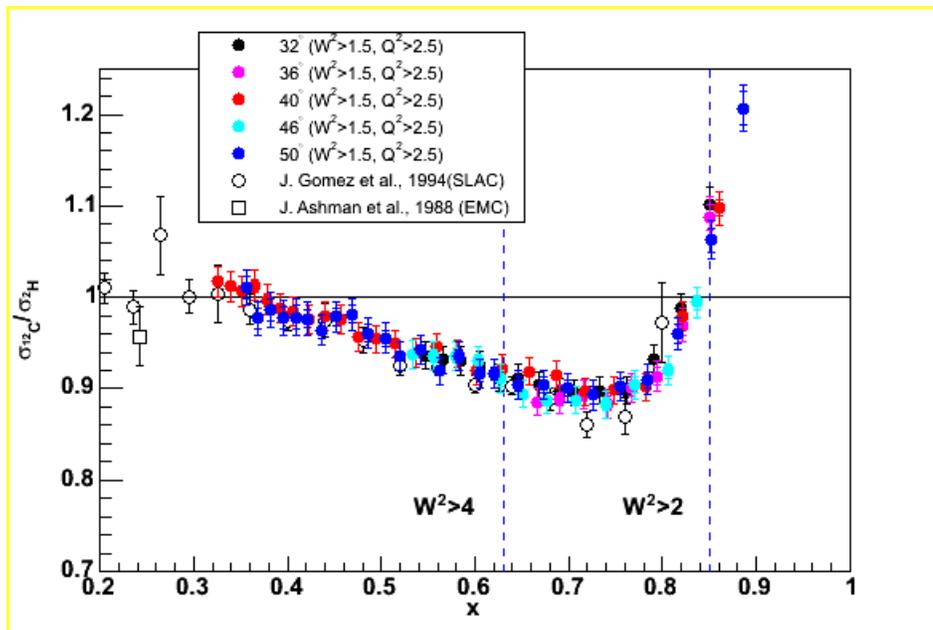}
\caption{Preliminary EMC ratios from E03-103 for $^{12}$C (top) for
the five largest $Q^2$ settings of the experiment.  The extracted EMC
ratios are consistent within the uncertainties for all settings, and there
is no systematic $Q^2$ dependence, up to $x=0.85$. Hollow points are data
from previous measurements~\cite{ashman88,gomez94}.}
\label{fig:qsqemc}
\end{center}
\end{figure}

A more precise test is made by examining the $Q^2$ dependence of the EMC
ratio, where many of the systematic uncertainties cancel in the ratio.
Figure~\ref{fig:qsqemc} shows the $Q^2$ dependence of the EMC ratio for carbon
for the five largest $Q^2$ values taken during the experiment - three at
5.8~GeV beam energy, and two (36$\deg$ and 46$\deg$) at 5.0~GeV.  The EMC
ratios are very precise, and are independent of $Q^2$ over the entire range of
$x$. While the $Q^2$ dependence of the structure functions is quite different
when taken at fixed $x$ as opposed to fixed $\xi$, the effect on the target
ratios is extremely small, and the EMC ratios are independent of $Q^2$ whether
taken as a function of $x$ or $\xi$.

\begin{figure}[ht]
\includegraphics[width=22pc]{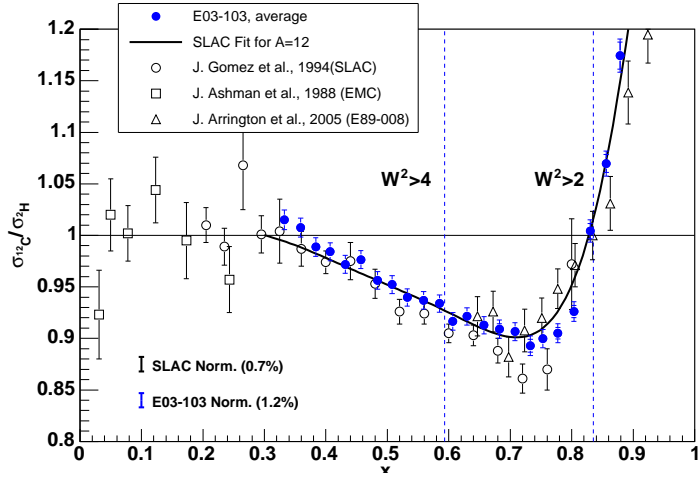}\hspace{2pc}\\
\includegraphics[width=22pc]{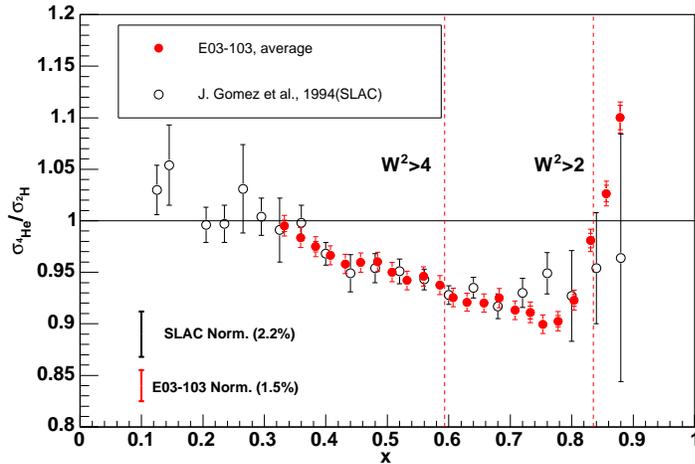}
\begin{minipage}[b]{14pc}\caption{Preliminary EMC ratios from E03-103 for the
$^{12}$C (top), $^4$He (bottom).  Inner error bars are statistical, while the
outer error bars are the combined statistical and estimated systematic
uncertainty. The hollow points are data from previous
measurements~\cite{ashman88,gomez94,arrington06a}.  Error bars in
the lower left corner of each plot indicate the estimated normalization
uncertainty of the previous and present measurements.\\}
\label{fig:emc}
\end{minipage}
\end{figure}

The detailed measurements of the $Q^2$ dependence were only taken for Carbon
and Deuterium. For the extraction of the EMC effect, we limit ourselves to the
two data sets at the largest $Q^2$ values.  These are the 40$\deg$ and 50$\deg$
settings taken with a beam energy of 5.8~GeV, where data was taken for all
nuclei.  Figure~\ref{fig:emc} shows the preliminary results for extracted EMC
ratios for $^{12}$C and $^4$He.  For these isoscalar nuclei, the cross section
ratios are identical to the usual EMC ratios, assuming that $R =
\sigma_L/\sigma_T$ is $A$-independent.  The $^{12}$C EMC ratio is consistent
with previous data, but of much higher precision at large $x$ values, where
previous measurements were statistics limited, mainly due to the large $Q^2$
values required to reach the DIS regime.  While the SLAC data suggested that
the EMC effect in $^4$He was slightly smaller than for $^{12}$C, these new
results indicate that the EMC is nearly identical to the $^{12}$C data. The
average density of $^4$He is anomalously large for such a light nucleus, and
is in fact nearly identical to the average nuclear density for $^{12}$C, so
these preliminary results are consistent with models where the EMC effect
scales with the nuclear density.

\begin{figure}[ht]
\includegraphics[width=22pc]{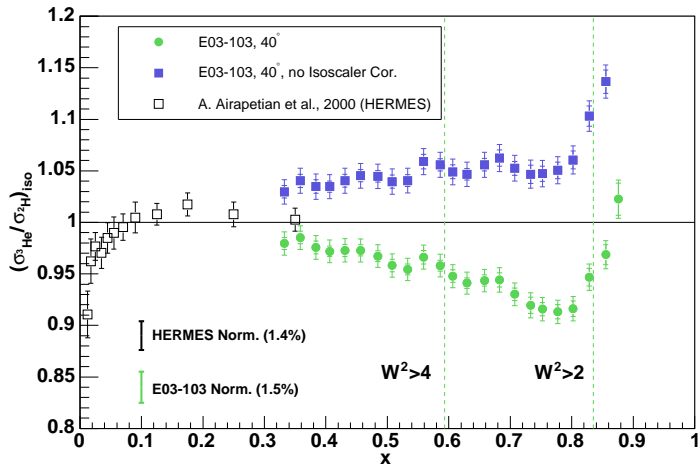}
\begin{minipage}[b]{14pc}\caption{Preliminary EMC ratios from E03-103 for the
$^3$He, errors are the same as in Fig.~\ref{fig:emc}.  The hollow points are
data from previous measurements~\cite{ackerstaff99}. The blue squares show the
raw $^3$He/$^2$H cross section ratio, while the green circles show the ratios
corrected to an isoscalar $A=3$ nucleus, taking the ratio of $\sigma_n /
\sigma_p$ from the SLAC fit used in Ref.~\cite{gomez94}.\\}
\label{fig:emc3}
\end{minipage}
\end{figure}

The results for $^3$He, shown in Figure~\ref{fig:emc3} do not provide the same
\textit{direct} measure of the EMC effect for $A=3$. Because $^3$He has two
protons and one neutron, we must apply a correction for the proton excess to
obtain the isoscalar EMC ratio, and there is uncertainty in the neutron
cross section in this region. The blue squares show the uncorrected cross
section ratio, while the green circles show the result after correcting for
the proton excess using the same fit as for SLAC E139: $\sigma_n / \sigma_p =
1 - 0.8x$.  The correction for $^3$He is nearly identical, but of the opposite
sign, to the correction applied to Au, which has a large neutron excess.  The
correction can be quite large, and the extracted EMC ratio can change by several
percent if one uses a different parameterization for the ratio $\sigma_n /
\sigma_p$. Thus, the interpretation of the $^3$He/$^2$H ratio is dependent on
the assumed isoscalar correction.  Because we also took data on $^1$H, 
we can extract the ratio of $^3$He to ($^1$H + $^2$H), which does not rely on
having a model for $\sigma_n / \sigma_p$.  Because comparing to the deuteron
takes into account a significant portion of the Fermi motion correction, which
is significant at a the largest $x$ values, the $x$ dependence of this ratio
will look quite different than the usual isoscalar corrected ratios.  However,
it can be more directly compared to calculations.

We are also examining extractions of the neutron cross section in this
region, and will be able to estimate the uncertainty in the $A=3$ EMC effect,
both by examining the uncertainties in $\sigma_n/\sigma_p$ and by examining
the $A$ dependence of the EMC effect. An isoscalar correction that
significantly increases the EMC effect for $A=3$ would have an equal but
opposite effect for Au, and so one can make a ``sanity check'' on the $A$
dependence resulting from different models of the isoscalar correction. In
addition, a consistency check can be performed by fitting the data for
$A$=1--4, if one has a model of both $\sigma_n/\sigma_p$ and the EMC effect. 
However, one cannot yet make an unambiguous separation between these two
effects. A new experiment~\cite{e03012} was recently completed at Jefferson
Lab which was designed to measure inclusive scattering from the neutron using
a deuteron target and tagging a slow, backwards proton to identify
events where the electron struck an effectively ``free'' neutron.  This
will provide further constraints on the correction to the $^3$He EMC ratios.

While there is a large uncertainty in the present extraction of the $A=3$
EMC effect, this preliminary data suggest that the EMC effect for $^3$He
may be larger than one would expect if one simply scales the EMC effect
by nuclear mass or average nuclear density.  In either of these cases, one
expects a much smaller effect for $A=3$ than observed in Carbon.  This
suggests the possibility that the EMC effect in the deuteron may also be larger
than expected.  However, additional study of the isoscalar correction is
needed before one can make any significant conclusions from the $^3$He
measurements.

\section{Conclusions}

We have presented preliminary results for the EMC effect in light nuclei.
The EMC effect for $^4$He is nearly identical to that for $^{12}$C, indicating
that the modification of the quark distribution scales with the average nuclear
density.  The data for $^3$He suggest that the EMC effect may also be large
for $A=3$ systems, but the interpretation depends on the model of the neutron
cross section used to form the ratio of isoscalar nuclei.

Results for the heavier nuclei will soon be available, which will allow us
to examine the $A$ dependence of the EMC effect at large $x$, as well as
examine the effect of the isoscalar correction on the $A$ dependence.  
In addition, we will be able to provide the ratio of $^3$He to the sum of
proton plus deuteron, avoiding the model dependence associated with the
neutron cross section correction.  Finally, we will provide a more detailed
examination of the scaling of the nuclear structure function, both in the
kinematic region discussed here, and for the $x>1$ region~\cite{e02019}.

\ack
This work was supported in part by the U.S. Department of Energy, Office of
Nuclear Physics, under contracts DE-AC02-06CH11357. The Thomas Jefferson
National Accelerator Facility is operated by the Southeastern Universities
Research Association under DOE contract DE-AC05-84ER40150.

\bigskip

\bibliography{GHP06}

\end{document}